\documentclass{new_tlp}
\usepackage{mathptmx}
\usepackage{todonotes}
\usepackage{url}
\usepackage{yfonts}

\def\ar{\leftarrow}
\def\lrar{\leftrightarrow}
\def\beq{\begin{equation}}
\def\eeq#1{\label{#1}\end{equation}}
\def\ba{\begin{array}}
\def\ea{\end{array}}
\def\gringo{{\sc gringo}}
\def\clingo{{\sc clingo}}
\def\anthem{{\sc anthem}}
\def\vampire{{\sc vampire}}
\def\num{\overline}
\def\p2f{\hbox{p2f}}

\newcommand{\I}{\mathcal{I}}
\newcommand{\J}{\mathcal{J}}
\newcommand{\M}{\mathcal{M}}
\newcommand{\PP}{\mathcal{P}}

\title[External Behavior and Verification of Refactoring]{External Behavior of a Logic Program\\ and Verification of Refactoring}

\author[J. Fandinno et al.]
  {JORGE FANDINNO, ZACHARY HANSEN, YULIYA LIERLER\\
  University of Nebraska Omaha
  \and VLADIMIR LIFSCHITZ, NATHAN TEMPLE\\
  University of Texas at Austin}

\jdate{March 2003}
\pubyear{2003}
\pagerange{\pageref{firstpage}--\pageref{lastpage}}
\doi{S1471068401001193}

\begin{document}

\label{firstpage}

\maketitle

\begin{abstract}
  Refactoring is modifying a program without changing its external behavior.
  In this paper, we
  make the concept of external behavior precise for a simple answer set
  programming language.  Then we describe a proof assistant for the
  task of verifying
  that refactoring a program in that language is performed correctly.
\end{abstract}

  
\section{Introduction}

This paper is about the process of refactoring in the context of answer set
programming~(ASP),
that is, about modifying an ASP program without changing
its external behavior.  Examples of refactoring logic programs can be
found in papers by~\citeANP{ser03} (\citeyearNP{ser03}), \citeANP{geb11b} (\citeyearNP{geb11b}, Section~3.1) and~\citeANP{bud15} 
(\citeyearNP{bud15}, Section~3).  In this paper we
propose, for a simple ASP language, a precise definition
of external behavior and a method for verifying that two programs
exhibit the same external behavior.

Refactoring a program usually involves a series of
small changes that improve its structure or performance.
The example below shows that, in ASP, refactoring may serve also
another purpose: to transform a program
that a grounder classifies as unsafe into an equivalent program that it
is able to ground.  The program
\begin{verbatim}
  composite(I*J) :- I > 1, J > 1.
        prime(I) :- I = a..b, not composite(I).
\end{verbatim}
defines the set of primes in the
interval $\{a,\dots,b\}$, assuming that $a>1$.  The grounder
{\sc gringo} \cite{gringomanual}
tells us that the program is unsafe. A safe program
defining the same set can be obtained by replacing the first rule with
\begin{verbatim}
  composite(I*J) :- I = 2..b, J = 2..b.
\end{verbatim}
This is an example of refactoring, because the extent of \verb|prime/1|
did not change.

We can also refactor the program to improve its performance
using the fact that every composite number in $\{a,\dots,b\}$ has a
divisor in the interval $\{2,\dots\lfloor \sqrt b\rfloor\}$:
\begin{verbatim}
       sqrt_b(M) :- M = 1..b, M*M <= b, (M+1)*(M+1) > b.
  composite(I*J) :- sqrt_b(M), I = 2..M, J = 2..b.
        prime(I) :- I = a..b, not composite(I).
\end{verbatim}

In the Abstract Gringo language \cite{geb15}, a program is defined as
a set of rules, so that a program
includes neither directives nor comments.  Under this narrow definition,
the program itself does not tell us which predicate symbols are meant to
represent the output, and which symbols are
auxiliary.  But this difference is essential, because changing
auxiliary predicates does not indicate a mistake in the process of
refactoring.

Furthermore, the rules of a program do not show what kind of input is
supposed to be provided for it.
Generally, an input for an ASP program can be specified in two ways.
First, some symbolic constants, such as {\tt a} and {\tt b} in the
programs above, may be meant to serve as placeholders for elements of
the input.  Second, some predicate symbols occurring in the program may occur
in the bodies of rules only, not in the heads.  The extents of such
predicates may be specified as part of input when we run the program.
Some inputs may not conform to the programmer's assumptions about
the intended use of the program.  For instance, when we run the prime number
programs above, the placeholders {\tt a} and {\tt b} are expected to be
replaced by integers; the cases when they are replaced by symbolic constants
are not related to external behavior if the programs are
used as intended.

To sum up, \emph{what we consider external behavior of a set of rules
depends on how these rules are meant to be used.}
In Sections~\ref{sec:extbe}--\ref{sec:fn},
we make this idea precise for the subset of
Abstract Gringo called mini-\gringo\ (\citeNP{fan20}, Section~2;
\citeNP{lif22a}, Sections~2,~3).  After that, we
describe the proof assistant {\sc anthem-p2p},\footnote{Available
  at \url{https://github.com/ZachJHansen/anthem-p2p}.} which uses
the theorem prover \hbox{\vampire} \cite{vor13}
 to verify that two mini-\gringo\ programs have the
 same external behavior.  
 This proof assistant is built on top of the system
 \anthem\ \cite{fan20}, whose focus is on the related and yet different
 task of confirming that an ASP program adheres to its specification.
 The prime number programs  above are used as a running example.
To make the paper more
 self-contained, we have reviewed some background material in
 Appendices~A--C.

\section{On the syntax of mini-\gringo}\label{sec:mg}

There are minor syntactic differences between
mini-\gringo\ and the input language of the grounder \gringo,
explained by the fact the former is designed for theoretical studies, and the
latter for actual programming.
For example, the definition of \verb|sqrt_b/1| in the introduction,
rewritten in the syntax of mini-\gringo, becomes
$$
\emph{sqrt\_b\/}(M) \ar M = \num  1\,..\,b \,\land\, M\times M \leq b
\,\land\, (M+\num 1)\times(M+\num 1) > b.
$$
Overlined symbols, such as $\num 1$, are ``numerals''---syntactic objects
representing integers.
In examples of rules and programs, we will
freely switch between the two styles.

In mini-\gringo,
\emph{precomputed terms} are numerals,
  symbolic constants, and the symbols \emph{inf}, \emph{sup}.
  We assume that a total order on precomputed terms is chosen, such that
  \emph{inf} is its least element, \emph{sup} is its greatest element, and,
 for all integers~$m$ and~$n$,   $\num m < \num n$ iff $m<n$.
A  \emph{precomputed atom} is an expression of the form $p({\bf t})$, where~$p$
  is a symbolic constant and~$\bf t$ is a tuple of precomputed terms.
  A \emph{predicate symbol} is a pair $p/n$, where~$p$ is a symbolic
  constant and~$n$ is a nonnegative integer.  About a rule or another
  syntactic expressions we say that it \emph{contains} $p/n$ if it
  contains an atom of the form $p(t_1,\dots,t_n)$.

\section{External behavior}\label{sec:extbe}

{\bf Definition 1}$\;$
A \emph{user guide} is a quadruple
\beq
(\emph{PH},\emph{In},\emph{Out},\emph{Dom\/})
\eeq{ug}
where
\begin{itemize}
\item \emph{PH} is a finite set of symbolic constants, called
  \emph{placeholders},
\item \emph{In} and \emph{Out} are disjoint finite sets of predicate
  symbols, called \emph{input symbols} and \emph{output symbols},
  and
\item \emph{Dom} is a set such that each of its elements is a pair $(v,\I)$,
  where
  \begin{itemize}
  \item [(i)]$v$ is a function that maps elements of~\emph{PH} to
      precomputed terms that do not belong to~\emph{PH}, and
    \item  [(ii)] $\I$ is a subset of the set of precomputed atoms that contain
      an input symbol and do not contain placeholders.
\end{itemize}
\end{itemize}

The set \emph{Dom} is the \emph{domain} of the user guide, and
pairs $(v,\I)$ satisfying conditions~(i) and~(ii) are called \emph{inputs}.
An input $(v,\I)$ represents a way to choose the values of
placeholders and the extents of input predicates: for every
placeholder~$c$, specify $v(c)$ as its value, and add the atoms~$\I$
to the rules of the program as facts.  If~$\Pi$ is a mini-\gringo\ program
then $v(\Pi)$ stands for the program obtained from~$\Pi$ by
replacing every occurrence of every constant~$c$ in the domain of~$v$
by~$v(c)$.
Using this notation, we can say that
choosing $(v,\I)$ as input for~$\Pi$ amounts to
replacing~$\Pi$ by the program $v(\Pi)\cup\I$.

To use a program in accordance with user guide~(\ref{ug})
means to run it for inputs that belong to \emph{Dom}.  The
inputs that do not belong to \emph{Dom} are not related to the
external behavior of the program when it is used as intended.

\medskip\noindent{\bf Example 1}$\;$ The intended use of the
programs discussed in the introduction can be
described by user guide~(\ref{ug}) with
$\emph{PH} = \{a,b\}$, $\emph{In} = \emptyset$,
$\emph{Out} = \{\emph{prime}/1\}$,
and with the domain consisting of the inputs $(v,\emptyset)$ such that
$v(a)$, $v(b)$ are numerals.  (We could
choose also to include the condition $v(b)\geq v(a)>\num 1$.)
This user guide will be denoted by $\emph{UG}_p$.

\medskip\noindent{\bf Example 2}$\;$ We would like to describe the meaning
of the word \emph{orphan} by a logic program \cite[Section~4.1.2]{gel14}.
The intended use of such a
program can be described by user guide~(\ref{ug}) with
$$\emph{PH} = \emptyset,\
\emph{In} = \{\emph{father/2},\emph{mother/2}, \emph{living/1}\},\
\emph{Out} = \{\emph{orphan/1}\},$$
and with the domain consisting of all inputs.    We will denote this
user guide by~$\emph{UG}_o$.  In the next two sections, we examine
two possible definitions of \hbox{\emph{orphan/1}}
and consider the question of their
equivalence with respect to $\emph{UG}_o$.
\medskip

User guides are closely related to lp-functions \cite[Section~2]{gel02},
and also to \hbox{io-programs} \cite[Section~5]{fan20}, reviewed
in~\ref{appc}.

    An \emph{output atom} of a user guide~\emph{UG} is a precomputed atom
    that contains an output symbol of~\emph{UG}.
    
    \medskip\noindent{\bf Definition 2}$\;$
    Let~$(v,\I)$ be an input in the domain
    of a user guide~\emph{UG}, and let~$\Pi$ be
a mini-\gringo\ program such that the heads of its rules do not
contain input symbols of~\emph{UG}.  The \emph{external behavior} of~$\Pi$
for the user guide~\emph{UG} and the input~$(v,\I)$ is the collection
of all sets that can be represented as the intersection of a stable model
of $v(\Pi)\cup\I$ with the set of output atoms of~\emph{UG}.

\medskip\noindent{\bf Example~1, continued}$\;$ If~$\Pi$ is one of the
three prime number programs from the introduction, and $(v,\I)$ is an
input in the domain of $\emph{UG}_p$, then the program $v(\Pi)\cup\I$
is $v(\Pi)$, and it has a unique stable model.  If~$v$ is defined by
the conditions
$v(a)=\num{10}$, $v(b)=\num{15}$,
then that stable model includes the atoms $\emph{prime}(\num{11})$,
$\emph{prime}(\num{13})$, and some atoms containing
$\emph{composite}/1$.  The external behavior of each of the programs
for this input is
$\{\{\emph{prime}(\num{11}),\emph{prime}(\num{13})\}\}$.
For the safe and optimized versions, this external behavior
can be calculated by instructing \clingo\ to
find all answers for the file obtained from the program by appending the
directives
\begin{verbatim}
  #const a = 10.  #const b = 15.  #show prime/1.
\end{verbatim}

\medskip\noindent{\bf Example~2, continued}$\;$ If~$\Pi$ is the program
\beq\ba l
\verb|parent_living(X) :- father(Y,X), living(Y).|\\
\verb|parent_living(X) :- mother(Y,X), living(Y).|\\
\verb|       orphan(X) :- living(X), not parent_living(X).|
\ea\eeq{orphan1}
and $(v,\I)$ is an input in the domain of $\emph{UG}_o$, then the
program $v(\Pi)\cup\I$ is $\Pi\cup\I$, and it has a unique stable model.
If~$\I$ is
\beq\ba l
\{\emph{father\/}(\emph{jacob},\emph{joseph}),
\emph{mother\/}(\emph{rachel},\emph{joseph}),\\
\quad \quad \emph{living}(\emph{jacob}),
\emph{living}(\emph{rachel}),
\emph{living}(\emph{joseph})\},
\ea\eeq{oi}
then that stable model includes the atoms
$\emph{orphan\/}(\emph{jacob})$, $\emph{orphan\/}(\emph{rachel})$,
and some atoms containing predicate symbols other than \emph{orphan/1}.
The external behavior of this program for $\emph{UG}_o$ and input~(\ref{oi})
is
\beq
\{\{\emph{orphan\/}(\emph{jacob}),\emph{orphan\/}(\emph{rachel})\}\}.
\eeq{ebor}
It can be calculated by instructing \clingo\ to
find all answers for the file obtained from
program~(\ref{orphan1}) by appending the facts
\begin{verbatim}
  father(jacob,joseph).
  mother(rachel,joseph).
  living(jacob). living(rachel). living(joseph).
\end{verbatim}
and the directive \verb|#show orphan/1|.

\medskip
In the special case when \emph{UG} has neither placeholders nor input symbols,
and its set of output symbols includes all predicate symbols occurring
in~$\Pi$, the external
behavior of~$\Pi$ with respect to~\emph{UG} and $(\emptyset,\emptyset)$
is the set of stable models of~$\Pi$.  In this sense, the concept of
external behavior is a generalization of the stable model semantics.

\section{Equivalence}

\noindent{\bf Definition 3}$\;$
Let~\emph{UG} be a user guide, and let~$\Pi_1$, $\Pi_2$ be mini-\gringo\
programs such that the heads of their rules do not contain input symbols
of~\emph{UG}.  We say that~$\Pi_1$ is \emph{equivalent to~$\Pi_2$ with
respect to}~\emph{UG} if, for every input $(v,\I)$ in the domain of~\emph{UG},
the external behavior of $\Pi_1$ for~\emph{UG} and~$(v,\I)$ is the
same as the external behavior of~$\Pi_2$.

\medskip\noindent{\bf Example~1, continued}$\;$ The three programs from the
introduction are equivalent to each other with respect to $\emph{UG}_p$.
As discussed in Section~\ref{sec:func}, this claim can be verified using
the automated reasoning tools {\sc anthem-p2p} and \vampire.

\medskip\noindent{\bf Example~2, continued}$\;$ Perhaps surprisingly,
the one-rule program
\beq\ba l
\verb|orphan(X) :- living(X), father(Y,X), mother(Z,X),|\\
\verb|             not living(Y), not living(Z).|
\ea\eeq{orphan2}
is not equivalent to~(\ref{orphan1}) with respect to $\emph{UG}_o$.
Indeed, the external behavior of this program with respect
to $\emph{UG}_o$ and input~(\ref{oi}) is $\{\emptyset\}$, which
is different from~(\ref{ebor}).  We will see
that {\sc anthem-p2p} can help us clarify the relationship between
programs~(\ref{orphan1}) and~(\ref{orphan2}).

\medskip
We understand \emph{refactoring} a mini-\gringo\ program with respect to a
  user guide~\emph{UG} as replacing it by a program that is equivalent
  to it with respect to~\emph{UG}.

This equivalence relation is essentially an
example of relativized uniform equivalence with projection
\cite{oet08}, except that the language discussed by~\citeANP{oet08} includes
neither arithmetic operations nor placeholders.
It is \emph{uniform} equivalence, because the programs are extended by
adding facts, rather than more complex rules; \emph{relativized},
because these
  facts~$\I$ are assumed to be atoms containing input symbols,
  not arbitrary atoms; \emph{with projection}, because
  we look at the output atoms in the stable model, not the entire model.

\section{Formal notation for user guides} \label{sec:fn}

To design software for verifying the equivalence of
programs with respect to a user guide, we need to
represent user guides in formal notation.   The format that we chose for
user guide files is similar to the format of
specification files, defined by~\citeANP{fan20}~\citeyear{fan20} within their work on the
system \anthem.
Placeholders
and input symbols are represented by \verb|input| statements, for instance:
\begin{verbatim}
  input: n.
  input: living/1, father/2, mother/2.
\end{verbatim}
Output symbols are represented by \verb|output| statements:
\begin{verbatim}
  output: prime/1.
\end{verbatim}
There can be several statements of both kinds in a user guide file, in any
order.

The question of representing the domain \emph{Dom} by a string of
characters is more difficult, because the domain is a set of inputs,
which is generally infinite.
Our approach is to define ``assumptions'' as sentences of an
appropriate first-order language, and characterize the domain by a
list of assumptions; an input belongs to the domain iff it
satisfies all assumptions on that list.

For any set~$\PP$ of predicate symbols, by $\sigma_0(\PP)$ we denote the
subsignature of the two-sorted signature~$\sigma_0$, described in \ref{appa},
in which the set of predicate symbols
is limited to the comparison symbols and the symbols from~$\PP$.
In this paper, an \emph{assumption} is a sentence over the
signature~$\sigma_0(\emph{In\/})$.
Besides \verb|input| and \verb|output| statements, a user guide
file may include one or more statements consisting of the word
\verb|assume| followed by an assumption.

To use assumptions as conditions on an input, we need to relate inputs
to interpretations in the sense of first-order logic.
If~$v$ is a function that maps elements of some set~\emph{PH} of
symbolic constants to symbolic constants, and~$\I$ is a subset of the set of
precomputed atoms that contain a predicate symbol from~$\PP$,
then there exists a unique interpretation~$I$ of~$\sigma_0(\PP)$ such that
\begin{itemize}
\item[(a)]
the domain of the sort \emph{general} in~$I$
  is the set of all precomputed terms;
\item[(b)]
the domain of the sort \emph{integer} in~$I$ is the set of all numerals;
\item[(c)] $I$ interprets every symbolic constant~$c$ in~\emph{PH}
  as~$v(c)$;
\item[(d)] $I$ interprets every precomputed term~$t$ that does not
  belong to~\emph{PH} as~$t$;
\item[(e)] $I$ interprets the symbols for arithmetic operations
  as usual in arithmetic;
\item[(f)] if $p/n$ is a predicate constant from~$\PP$, and $\bf c$
  is an $n$-tuple of precomputed atoms, then $I$ interprets
  $p({\bf c})$ as true iff $p({\bf c})\in\I$;
\item[(g)] $I$ interprets the comparison symbols as in the definition of
  mini-\gringo.
\end{itemize}
We will denote that interpretation by~$I(v,\I)$.
The domain of the user guide defined by a set of assumptions is the set
of inputs $(v,\I)$ such that the interpretation $I(v,\I)$
of~$\sigma_0(\emph{In})$ satisfies all assumptions in that set.

\medskip\noindent{\bf Example~1, continued}$\;$ The user guide $\emph{UG}_p$
can be described by the statements
\begin{verbatim}
  input: a, b.
  assume: exists N (a = N) and exists N (b = N).
  output: prime/1.
\end{verbatim}
The first two lines can be written more concisely as
\begin{verbatim}
  input: a -> integer, b -> integer.
\end{verbatim}

\medskip\noindent{\bf Example~2, continued}$\;$ The user guide $\emph{UG}_o$
can be described by the statements
\beq\ba l
\verb| input: living/1, father/2, mother/2.|\\
\verb| output: orphan/1.|
\ea\eeq{ugo}
The absence of \verb|assume| statements here shows that the domain
is the set of all inputs.

\section{Functionality of {\sc anthem-p2p}} \label{sec:func}

The proof assistant {\sc anthem-p2p} uses the theorem prover \hbox{\vampire}
to verify that two mini-\gringo\ programs have the
same external behavior with respect to a given user guide.
We can verify, for instance,
that the first two versions of the prime number program
from the introduction are equivalent with respect to the user guide
$\emph{UG}_p$ by running {\sc anthem-p2p} on three files: the
unsafe program
\beq\ba l
\verb|composite(I*J) :- I > 1, J > 1.|\\
\verb|      prime(I) :- I = a..b, not composite(I).|
\ea\eeq{primes.1}
the safe program
\beq\ba l
\verb|composite(I*J) :- I = 2..b, J = 2..b.|\\
\verb|      prime(I) :- I = a..b, not composite(I).|
\ea\eeq{primes.2}
and the user guide
\beq\ba l
\verb|input: a -> integer, b -> integer.|\\
\verb|output: prime/1.|
\ea\eeq{primes.ug}

The system {\sc anthem-p2p} transforms the task of verifying equivalence
with respect to a user guide~(\ref{ug}) into the problem of verifying
the provability of a formula in a first-order theory over the
signature~$\sigma_0(\emph{In}\cup\emph{Out})$, and submits that problem
to \vampire; see Sections~\ref{sec:tight}--\ref{sec:ontop} for details.

The user can help \vampire\ organize search more efficiently by
supplying {\sc anthem-p2p} with ``helper'' files.  Such a file may instruct
\vampire\ to prove a series of lemmas before trying to prove the goal formula.
A helper file can suggest also instances of the induction schema that may be
useful for the job in hand.  This kind of help is needed, for instance, for
verifying the equivalence of the optimized prime number program to the
other two.

The use of {\sc anthem-p2p} for proving
equivalence of programs is, generally, an interactive process.
If \vampire\ does not prove the goal formula in the allotted time then one of
the options is to provide more lemmas and run {\sc anthem-p2p}
again.  Alternatively, the user can look for a counterexample that refutes the
equivalence claim, as in Example~2 above.

Sometimes, {\sc anthem-p2p} can 
help us clarify the source of a puzzling discrepancy between two versions
of a program if we run it in the presence of additional \verb|assume|
statements.  If
adding an assumption to the user guide makes the programs equivalent then
it is possible that perceiving that assumption as self-evident
is the reason why the discrepancy is puzzling.  For
instance, we can observe that
the {\sc anthem-p2p}/\vampire\ combination
proves the equivalence of
program~(\ref{orphan1}) to program~(\ref{orphan2}) if we extend user
guide~(\ref{ugo}) by two existence and uniqueness assumptions:
\begin{verbatim}
  assume: forall X exists Y forall Z (father(Z,X) <-> Y=Z).
  assume: forall X exists Y forall Z (mother(Z,X) <-> Y=Z).
\end{verbatim}

The limitations of the {\sc anthem-p2p} algorithm are inherited from the
limitations of \anthem\ and can be described as follows.
The \emph{predicate dependency graph} of a mini-\gringo\
program~$\Pi$ \cite[Section~6.3]{fan20} is the directed graph that
\begin{itemize}
\item has the predicate symbols contained in~$\Pi$ as its vertices, and
\item has an edge from~$p/n$ to~$q/m$ if some rule of~$\Pi$
  contains~$p/n$ in the head and $q/m$ in the body.
\end{itemize}
The edge from $p/n$ to $q/m$ is \emph{positive} if there is a rule~$R$
in~$\Pi$ such that~$p/n$ is contained in the head of~$R$, and $q/m$ is
contained in an atom in the body of~$R$ that is not in the scope of
negation.  For example, the predicate dependency graph
of program~(\ref{orphan1}) has~6 edges; all of them except for the edge
from \verb|parent_living/1| to \verb|orphan/1| are positive.
We say that~$\Pi$ is \emph{tight} if this graph has no
cycles consisting of positive edges.

A vertex~$p/n$ of the graph is \emph{private} for a user
guide~\emph{UG} if it is neither an input symbols nor an output
symbol of~\emph{UG}.  We say that~$\Pi$ \emph{uses private recursion}
for~\emph{UG} if
\begin{itemize}
\item the predicate dependency graph of~$\Pi$ has a cycle such that
  every vertex in it is a private symbol, or
\item $\Pi$ includes a choice rule with the head containing a private
  symbol.
\end{itemize}

As discussed in the next two sections, the
applicability of the algorithm implemented in {\sc anthem-p2p} to a pair
of mini-\gringo\ programs and a user guide~\emph{UG} is guaranteed whenever
the programs are tight and do not use private recursion with respect
to~\emph{UG}.  We expect that it will be possible to replace the tightness
requirement by a significantly weaker condition using the ideas of a
recent paper on ``locally tight'' programs \cite{fan21a}; this is a topic for
future work.

\section{Equivalence of tight programs}\label{sec:tight}

The theorem stated below relates equivalence of tight programs to the
satisfaction relation of second-order logic.  Its statement refers to
the concept of second-order completion, reviewed in \ref{appb},
and also to the concept of  standard interpretation, defined as follows.
An interpretation~$I$ of~$\sigma_0(\PP)$ is \emph{standard} for a
set~\emph{PH} of symbolic constants if it
satisfies conditions~(a), (b), (d), (e), (g)
from Section~\ref{sec:fn} and the condition
\begin{itemize}
\item[(c$'$)] $I$ interprets every symbolic constant in~\emph{PH}
  as a term that does not belong to~\emph{PH}.
\end{itemize}

\medskip\noindent{\bf Theorem}$\;$\emph{
  Let~UG be a user guide (PH,In,Out,Dom) such that its domain is described by
  a finite set of assumptions, and let Asm be the conjunction of these
  assumptions.
  For any tight mini-\gringo\ programs $\Pi_1$,~$\Pi_2$ such that the
  heads of their rules do not contain the input symbols of~UG,
$\Pi_1$ is equivalent to~$\Pi_2$ with respect to UG iff the sentence}
\beq
\emph{Asm}\to(\hbox{COMP}(\Pi_1,\emph{In},\emph{Out\/}) \lrar
\hbox{COMP}(\Pi_2,\emph{In},\emph{Out\/}))
\eeq{thmf}
\emph{is satisfied by all interpretations of the
  signature~$\sigma_0$(In\,$\cup$Out) that are standard for~PH.}
\medskip

This theorem shows that the equivalence of tight programs may be established
by choosing a first-order theory~$T$ over the
signature~$\sigma_0(\emph{In\/}\cup\emph{Out})$ such that its axioms are
satisfied by all interpretations that are
standard for~\emph{PH}, and then exhibiting a derivation of
formula~(\ref{thmf}) from the axioms of~$T$ in classical second-order logic.
For programs that do not use private recursion, the problem of constructing
such a derivation can be reduced to proof search in first-order logic
(see Section~\ref{sec:elim}
below), for which many automated reasoning tools are available.
This is the core of the procedure used by {\sc anthem-p2p}.

The proof of the theorem, including the lemma below, uses
terminology related to io-programs, which
is reviewed in \ref{appc}.

\medskip\noindent{\bf Lemma}$\;$\emph{
Let~$\Pi$ be a mini-\gringo\ program such that the heads of its rules do not
contain input symbols of a user guide}
(\emph{PH},\emph{In},\emph{Out},\emph{Dom\/}). \emph{For any
  input~$(v,\I)$, a set~$\J$ of
  output atoms is an element of the external behavior
  of~$\Pi$ for} (\emph{PH},\emph{In},\emph{Out},\emph{Dom\/})
\emph{and $(v,\I)$ iff
$\I\cup\J$ is an io-model of the io-program}
($\Pi$,\emph{PH},\emph{In},\emph{Out}) \emph{for $(v,\I)$.}

\medskip\noindent{\bf Proof}$\;$
For every set~$\J$ of output atoms, the conditions
\begin{itemize}
\item $\J$ is the set of all output atoms in some stable model~$\M$ 
  of $v(\Pi)\cup\I$;
\item $\I\cup\J$ is the set of all public atoms in some stable
  model~$\M$ of $v(\Pi)\cup\I$
\end{itemize}
are equivalent to each other.  Indeed,
since the heads of rules of~$v(\Pi)$ do not contain
input atoms, the set of input atoms in~$\M$ is~$\I$.

\medskip\noindent{\bf Proof of the Theorem}$\;$
The condition
\beq
\Pi_1\hbox{ is equivalent to }\Pi_2\hbox{ with respect to }\emph{UG}
\eeq{pt1}
means that for any input $(v,\I)$ such that $I(v,\I)\models\emph{Asm}$
and any set~$\J$ of output atoms,
\beq\ba c
\J\hbox{ is an element of the external behavior
  of~$\Pi_1$ for \emph{UG} and }(v,\I)\\
\hbox{iff}\\
\J\hbox{ is an element of the external behavior
  of~$\Pi_2$ for \emph{UG} and }(v,\I).
\ea\eeq{pt2}
By the lemma, condition~(\ref{pt2}) can be reformulated as follows:
$$\ba c
\I\cup\J\hbox{ is an io-model of the io-program
($\Pi_1$,\emph{PH},\emph{In},\emph{Out}) for }(v,\I)\\
\hbox{iff}\\
\I\cup\J\hbox{ is an io-model of the io-program
($\Pi_2$,\emph{PH},\emph{In},\emph{Out}) for }(v,\I).\\
\ea$$
By the theorem quoted at the end of \ref{appc}, this can be further
reformulated as
\beq I(v,\I\cup\J) \models
\hbox{COMP($\Pi_1$,\emph{In\/},\emph{Out\/})}
\lrar
\hbox{COMP($\Pi_2$,\emph{In\/},\emph{Out\/})}.
\eeq{pt3}
Hence condition~(\ref{pt1}) is equivalent to requiring that~(\ref{pt3})
hold for all inputs $(v,\I)$ such that
$I(v,\I)\models\emph{Asm}$ and all set~$\J$ of output atoms.

Since assumptions do not contain output symbols,
$I(v,\I)\models\emph{Asm}$ is equivalent to $I(v,\I\cup\J)\models\emph{Asm}$.
It follows that~(\ref{pt1}) is equivalent to asserting that
implication~(\ref{thmf}) is satisfied by $I(v,\I\cup\J)$ for all
inputs $(v,\I)$ and all sets~$\J$ of output atoms.  It remains to observe
that an interpretation of the
signature~$\sigma_0(\emph{In\/}\cup\emph{Out\/})$ can be represented in
the form $I(v,\I\cup\J)$ if and only if it is
standard for~\emph{PH}.

\section{Reduction to first-order logic} \label{sec:elim}
If $\Pi_1$ and $\Pi_2$ do not use private recursion then
the reference to second-order consequences of the
axioms of~$T$ in Section~\ref{sec:tight} can be eliminated in the
following way.  Represent the formula
COMP$(\Pi_1,\emph{In\/},\emph{Out\/})$ in the form
$$
\exists {\bf P}\left(\bigwedge_i F_i({\bf P}) \land F'({\bf P})\right),
$$
where $\bf P$ is a list
of distinct predicate variables
corresponding to the
private symbols $p_1,p_2,\dots$
of $\Pi_1$, and~$F_i({\bf P})$ is
the formula obtained from the completed definition
of~$p_i$ in~$\Pi_1$ by replacing each of $p_1,p_2,\dots$ by the
corresponding member of~$\bf P$. (Thus the conjunctive members
of $F'({\bf P})$ correspond to the completed definitions of the output
symbols and to the constraints of~$\Pi_1$.)  Similarly, write
COMP$(\Pi_2,\emph{In\/},\emph{Out\/})$ as
\beq
\exists {\bf Q}\left(\bigwedge_j G_j({\bf Q}) \land G'({\bf Q})\right),
\eeq{comp2}
where $\bf Q$ is a list of distinct predicate variables corresponding to the
private symbols $q_1,q_2,\dots$ of~$\Pi_2$,
and the formulas $G_j({\bf Q})$ are obtained from the completed
definitions of these symbols in~$\Pi_2$ by replacing them with
corresponding variables.  Take one half
\beq
\emph{Asm}\to(\hbox{COMP}(\Pi_1,\emph{In},\emph{Out\/}) \to
\hbox{COMP}(\Pi_2,\emph{In},\emph{Out\/}))
\eeq{thmf1}
of condition~(\ref{thmf}).
Since~$\Pi_2$ does not use private
recursion, formula~(\ref{comp2}) is equivalent to
$$\forall {\bf Q}\left(\bigwedge_j G_j({\bf Q}) \to G'({\bf Q})\right)$$
\cite[Theorem~3]{fan20}.  It follows that formula~(\ref{thmf1}) is
equivalent to
$$
\emph{Asm}\to
 \left(\exists {\bf P}\left(\bigwedge_i F_i({\bf P}) \land F'({\bf P})\right)
 \to\forall {\bf Q}\left(\bigwedge_j G_j({\bf Q}) \to G'({\bf Q})\right)\right)
$$
and consequently to
\beq
\forall {\bf PQ}\left(\left(
    \emph{Asm}\land\bigwedge_i F_i({\bf P}) 
\land\bigwedge_j G_j({\bf Q}) \right) \to (F'({\bf P})\to G'({\bf Q}))\right)
\eeq{onehalf}
(with the bound variables in {\bf P}, {\bf Q} renamed, if necessary, to ensure
that they are pairwise disjoint).
Similarly, the second half
$$
\emph{Asm}\to(\hbox{COMP}(\Pi_2,\emph{In},\emph{Out\/}) \to
\hbox{COMP}(\Pi_1,\emph{In},\emph{Out\/}))
$$
of condition~(\ref{thmf}) is equivalent to the formula obtained
from~(\ref{onehalf}) by swapping $F'({\bf P})$ with $G'({\bf Q})$.
Thus~(\ref{thmf}) can be rewritten as
$$
\forall {\bf PQ}\left(\left(
   \emph{Asm}\land\bigwedge_i F_i({\bf P}) 
   \land\bigwedge_j G_j({\bf Q}) \right) \to
          (F'({\bf P})\lrar G'({\bf Q}))\right).
$$
Finally, observe that this formula is entailed by the axioms of~$T$ if
and only if the axioms entail the first-order formula
\beq
\left(
    \emph{Asm}\land\bigwedge_i F_i({\bf p})
\land\bigwedge_j G_j({\bf q}) \right) \to (F'({\bf p})\lrar G'({\bf q})),
\eeq{thmffo}
where {\bf p}, {\bf q} are lists of fresh predicate constants.

We return to this formula in the description of the design of
{\sc anthem-p2p} below.  Note that its subformulas
$F_i({\bf p})$, $G_j({\bf q})$,
$F'({\bf p})$,
$G'({\bf q})$ are parts of the first-order completion
formulas of~$\Pi_1$ and~$\Pi_2$, modified by replacing their private symbols
$p_1,p_2,\dots$, $q_1,q_2,\dots$ by members of the lists~{\bf p} and~{\bf q}.

\section{Design of {\sc anthem-p2p}} \label{sec:ontop}

The system {\sc anthem-p2p}
is a Python program than operates by converting a claim
about the equivalence of two mini-\gringo\ programs
into an input for \anthem.
The system \anthem\ verifies the correctness of an io-program with
respect to a formal specification.  The file describing a specification
includes lists of placeholders, input symbols, output symbols, and
assumptions, and also a list of ``specs''
that describe the intended behavior of the future program by sentences over the
signature~$\sigma_0(\emph{In\/}\cup\emph{Out\/})$.

Given programs~$\Pi_1$ and~$\Pi_2$ and a user guide
(\emph{PH},\emph{In},\emph{Out},\emph{Dom\/}) with the domain described
by assumptions~\emph{Asm}, {\sc anthem-p2p} constructs the following
specification~\emph{Sp\/}:
\begin{itemize}
\item[(i)]
  the placeholders of~\emph{Sp\/} are the placeholders~\emph{PH\/} of the
  given user guide;
  \item[(ii)]
    the input symbols of~\emph{Sp\/} are the input symbols~\emph{In} of
    the user guide and the predicate symbols~{\bf p} corresponding to
    the private symbols $p_1,p_2,\dots$ of the program $\Pi_1$;
  \item[(iii)]
    the output symbols of~\emph{Sp\/} are the output symbols~\emph{Out} of
the user guide;
\item[(iv)]
  the assumptions of~\emph{Sp\/} are the assumptions~\emph{Asm} of the
  user guide and the modified completed definitions~$F_i({\bf p})$ of the
  private symbols of~$\Pi_1$;
\item[(v)]
  the specs of~\emph{Sp\/} are the remaining conjunctive terms
  $F'({\bf p})$ of the modified first-order completion formula of~$\Pi_1$.
\end{itemize}
Then {\sc anthem-p2p} 
instructs \anthem\ to prove the claim
that the io-program $(\Pi_2,\emph{PH},\emph{In},\emph{Out})$
implements~\emph{Sp\/}.
Providing \anthem\ with such an instruction makes it to look for a derivation of the formula
\beq\left(
    \emph{Asm}\land\bigwedge_i F_i({\bf p})
\land\bigwedge_j G_j({\bf q}) \right) \to (G'({\bf q})\lrar F'({\bf p}))
\eeq{goal}
from the axioms of~$T$ by invoking the theorem prover
\vampire~\cite[Section~6.4]{fan20}.
This formula is equivalent to~(\ref{thmffo}).
Thus instructing \anthem\ to verify that
the io-program $(\Pi_2,\emph{PH},\emph{In},\emph{Out})$
implements the specification~\emph{Sp\/} amounts to verifying the provability
of formula~(\ref{thmffo}) in~$T$. 

As an example, consider the operation of the {\sc anthem-p2p} algorithm on
programs~(\ref{primes.1}) and~(\ref{primes.2}) and user
guide~(\ref{primes.ug}).  In each of the  programs, the only private
predicate is \verb|composite/1|; it corresponds to both $p_1$ and~$q_1$ in the
notation of Section~\ref{sec:elim}.  The symbols \verb|composite_1/1|
and \verb|composite_2/1|, generated by {\sc anthem-p2p},
play the parts of~$\bf p$ and~$\bf q$ in formula~(\ref{thmffo}).  The
file describing the specification~\emph{Sp\/} is obtained in this case from
user guide~(\ref{primes.ug}) by adding three statements.  First, in
accordance with clause~(ii) in the description of \emph{Sp\/} above,
{\sc anthem-p2p} adds the statement
\begin{verbatim}
  input: composite_1/1.
\end{verbatim}
Second, in accordance with clause~(iv), a definition of
\verb|composite_1/1| is assumed:
\begin{verbatim}
  assume: forall X (composite_1(X) <->
                    exists N1, N2 (N1 > 1 and N2 > 1 and X = N1 * N2)).
\end{verbatim}
Finally, in accordance with clause~(v), a definition of \verb|prime/1|
in terms of \verb|composite_1/1| is added as a spec:
\begin{verbatim}
  spec: forall X (prime(X) <->
                  exists N1 (not composite_1(N1) and
                      exists N2, N3 (N2 = a and N3 = b and
                                     N2 <= N1 and N1 <= N3) and X = N1)).
\end{verbatim}

Once~\emph{Sp} is generated, \anthem\ calls
\vampire\ to prove formula~(\ref{goal}) in the theory~$T$, first by deriving
the specs~$F'({\bf p})$ from the antecedent of~(\ref{goal}) and $G'({\bf q})$
(``verification of specification from translated program''), and then by
deriving $G'({\bf q})$ from  the antecedent of~(\ref{goal})
and the specs~$F'({\bf p})$ (``verification
of translated program from specification'').  In this example, the
runtime of \vampire\ will be significantly reduced (a few seconds instead
of a few minutes) if we instruct it to start by proving two lemmas:
\begin{verbatim}
  lemma: forall I, J (I > 1 and J > 1 -> I < I*J).
  lemma: forall X (prime(X) ->
                   exists N1 (not composite_1(N1) and
                       exists N2, N3 (N2 = a and N3 = b and
                                      N2 <= N1 and N1 <= N3) and X = N1)).
\end{verbatim}

The {\sc anthem-p2p} website\footnote{Available at \url{https://ap2p.unomaha.edu/}.} allows users to 
experiment with the system in their web browser. 
The proof search is conducted on a University of Nebraska Omaha
server (Oracle Linux 8, 4 Intel(R) Xeon(R) Gold 6248 CPUs, 4 GB RAM) subject to a 10 minute timeout.
For smaller problems, this is the recommended introduction to the system.

\section{Conclusion}

This paper contributes to the theory of logic programming by defining user
guides, external behaviors, and equivalence with respect to a user guide.
The theorem proved in Section~\ref{sec:tight} relates equivalence of tight
programs to program completion.

The problem of checking equivalence between programs arises in many areas
of computer science.  For example, verifying the correctness of the
translation performed by an optimizing compiler is a problem of
this kind.  What is special about the verification of refactoring is that it
involves a pair of similar programs written in the same programming language.
{\sc mediator} \cite{dil18} is a tool that uses an SMT solver for the
verification of database refactoring.

The proof assistant {\sc anthem-p2p} can be used for verifying the
correctness of refactoring an ASP program, and also for comparing alternative
solutions to the same programming problem (for instance, in classroom
teaching and in ASP programming contests).
To make this tool more versatile, we plan to make it applicable to
programs with aggregates, along the lines of recent publications
(\citeNP{fan22a,lif22a}).

\section*{Acknowledgements}

We are grateful to Isil Dillig, Warren Hunt and
Jayadev Misra for valuable comments on the problems discussed in this paper.
Thanks also to Michael Gelfond and to the anonymous referees for advice
on improving the previous version.

\appendix

\section{Two-sorted formulas} \label{appa}

The signature~$\sigma_0$
has two sorts: the sort \emph{general} and its subsort
\emph{integer}.
Variables of the first sort are meant to
range over arbitrary precomputed terms, and we will identify them with
variables used in mini-\gringo\ rules.  Variables of the second sort are meant
to range over numerals---or, equivalently, integers.
The signature includes
\begin{itemize}
\item all precomputed terms as object constants; an object constant
  is assigned the sort \emph{integer} iff it is a numeral;
\item the symbols~$+$, $-$ and~$\times$ as binary function constants;
  their arguments and values have the sort \emph{integer};
\item all predicate symbols~$p/n$ as $n$-ary predicate
  constants; their arguments have the sort \emph{general};
\item the comparison symbols
$\neq\ <\ >\ \leq\ \geq$
as binary predicate constants; their arguments have the sort \emph{general}.
\end{itemize}
An atomic formula $(p/n)({\bf t})$ can be abbreviated
as $p({\bf t})$. An atomic formula $\prec\!\!(t_1,t_2)$, where~$\prec$ is a
comparison symbol, can be written as $t_1\prec t_2$.

In this paper, we adopt the convention that 
general variables start with $U$, $V$, $W$, $X$, $Y$, and $Z$;
integer variables start with $I$, $J$, $K$, $L$, $M$, and $N$.
For example, the formula
$\exists X (N=X)$ expresses that the value of~$N$ is an object of the sort
\emph{general\/};  it is universally true, because \emph{integer} is a
subsort of \emph{general}.  The formula
$\exists N (N=X)$ expresses that the value of~$X$ is an object of the sort
\emph{integer\/}; it is generally not true.

\section{Second-order completion} \label{appb}

Second-order completion \cite[Sections~6.1,~6.2]{fan20}
is a generalization of Clark's completion
\cite{cla78}
that uses bound predicate variables to model auxiliary
(``private'') predicates, such as \verb|composite/1| in our prime number
programs.  The definition covering the full
syntax of mini-\gringo\ is rather lengthy, and in this appendix we only give
an outline and an example.

Let \emph{In} and \emph{Out} be disjoint sets of predicate symbols, and
let~$\Pi$ be a mini-\gringo\ program such that atoms in the heads of
its rules do not contain predicate symbols from~\emph{In}.
If a predicate symbol~$p/n$
\begin{itemize}
\item is contained in an atom that occurs in a rule of~$\Pi$, and
\item belongs neither to~\emph{In} nor to~\emph{Out},
\end{itemize}
then~$p/n$ is a \emph{private symbol} of~$\Pi$.  We denote the
set of private symbols of~$\Pi$ by~\emph{Prv}.

The \emph{first-order completion} of~$\Pi$ is the conjunction of the following
first-order sentences over the
signature~$\sigma_0(\emph{In\/}\cup\emph{Out\/}\cup\emph{Prv\/})$:
\begin{itemize}
\item the completed definitions of the predicate symbols
  from~$\emph{Out\/}\cup\emph{Prv\/}$ in~$\Pi$;
\item the constraints of~$\Pi$ rewritten in the syntax of first-order
  logic.
\end{itemize}
The \emph{second-order completion} of~$\Pi$ is the sentence
over the signature $\sigma_0(\emph{In\/}\cup\emph{Out\/})$ obtained from
the first-order completion of~$\Pi$ by replacing all private symbols by
predicate variables and binding these variables by an existential quantifier.
We will denote the second-order completion of~$\Pi$ by
COMP$(\Pi,\emph{In\/},\emph{Out\/})$.

If, for instance, $\emph{In}=\emptyset$, $\emph{Out}=\{q/2\}$, and~$\Pi$
is the program
$$
\ba l
p(a),\\
p(b),\\
q(X,Y) \ar p(X) \land p(Y),
\ea$$
then $\emph{Prv}=\{p/1\}$,
the first-order completion of~$\Pi$ is
$$\ba l
\forall V(p(V) \lrar V=a \lor V=b)\,\land\\
\forall V_1V_2(q(V_1,V_2) \lrar \exists XY(q(V_1,V_2) \land p(X) \land p(Y)
\land V_1=X \land V_2=Y)),
\ea$$
and COMP$(\Pi,\emph{In\/},\emph{Out\/})$ is
$$
\ba l
\exists P(\forall V(P(V) \lrar V=a \lor V=b)\,\land\\
\forall V_1V_2(q(V_1,V_2) \lrar \exists XY(q(V_1,V_2) \land
P(X) \land P(Y) \land V_1=X \land V_2=Y))).
\ea
$$
This formula is equivalent to the first-order sentence
$$
\forall V_1V_2(q(V_1,V_2) \lrar (V_1=a \lor V_1=b)\land(V_2=a \lor V_2=b)).
$$

\section{Programs with input and output} \label{appc}

A \emph{program with input and output}, or an \emph{io-program},
is a quadruple
\beq
(\Pi,\emph{PH},\emph{In},\emph{Out\/}),
\eeq{iop}
where \emph{PH}, \emph{In} and \emph{Out} are as in the definition of
a user guide (Section~\ref{sec:extbe}), and~$\Pi$ is a mini-\gringo\ program
such that the heads of its rules do not contain symbols from \emph{In}
\cite[Section~5.1]{fan20}.  Inputs for an io-program are defined in the same
way as inputs for a user guide in Section~\ref{sec:extbe}.

A \emph{public atom} of an io-program~(\ref{iop}) is a precomputed atom
that contains a predicate symbol from $\emph{In\/}\cup\emph{Out\/}$.

An \emph{io-model} of an io-program~(\ref{iop}) for an input $(v,\I)$ is
a set that can be represented as the intersection of a stable model
of $v(\Pi)\cup\I$ with the set of public atoms of~(\ref{iop}).

If~$(v,\I)$ is an input for an io-program~(\ref{iop}), and the program~$\Pi$
is tight, then, for any set~$\J$ of output atoms, $\I\cup\J$ is an io-model
of~(\ref{iop}) iff the interpretation $I(v,\I\cup\J)$ of the
signature~$\sigma_0(\emph{In\/}\cup\emph{Out\/})$ satisfies the
second-order completion sentence COMP$(\Pi,\emph{In\/},\emph{Out\/})$
\cite[Theorem~2]{fan20}.

\bibliographystyle{acmtrans}
\bibliography{bib}

\end{document}